# Can future reirradiation justify upfront proton therapy? An extension of model-based proton therapy selection beyond primary-course NTCP benefit

*Nathan Torelli, Jan Unkelbach*

*Department of Radiation Oncology, University Hospital Zurich and University of Zurich, Zurich, Switzerland*

**Abstract**

**Objective:** Current model-based proton therapy selection values normal tissue sparing only through its immediate reduction in normal tissue complication probability (NTCP). If reirradiation later adds further dose to the normal tissue, sparing achieved during primary treatment may become more clinically valuable. We investigated whether, and under which conditions, this contingent benefit could justify upfront proton therapy.

**Approach:** We formulated a reirradiation-aware criterion that evaluates the expected NTCP benefit of upfront proton therapy over two scenarios: primary radiotherapy alone or primary radiotherapy followed by reirradiation, with the scenario-specific benefits weighted by their respective probabilities. Proton therapy is recommended when this expected benefit reaches a referral threshold $T$. For the analytically tractable case in which the toxicity outcomes of both scenarios could be represented by a common logistic NTCP model, we derived screening conditions based only on the paired primary treatments to determine whether an otherwise not-referred patient could become eligible under at least one future reirradiation scenario. We then quantified the minimum reirradiation probability and admissible added risk-score contributions required for additional referral, and mapped referral patterns across primary risk-score profiles and reirradiation scenarios.

**Main results:** Additional referral occurred for patients with substantial primary normal tissue sparing but a small immediate NTCP reduction because both plans lay on a flat region of the NTCP curve. In these cases, reirradiation could add sufficient normal tissue dose for the retained sparing to yield a larger NTCP benefit. Higher reirradiation probability expanded eligibility. Instead, higher reirradiation dose could reduce the NTCP benefit when both predictions approached saturation.

**Significance:** The proposed framework identifies well-defined patient profiles for which upfront proton therapy may be justified even if the NTCP benefit of proton therapy in the first treatment is below the threshold $T$. Clinical implementation will require validated reirradiation outcome models and individualized estimates of reirradiation probability.

## 1. Introduction

Proton therapy can substantially reduce radiation exposure to normal tissues compared with photon therapy, but its limited availability and higher cost require careful patient selection (van der Laan *et al*, 2012; Jakobi *et al*, 2015; Verma *et al*, 2016). In countries such as the Netherlands and Denmark, proton therapy is allocated using a model-based approach: photon and proton plans are generated for the same patient, their normal tissue complication probabilities (NTCP) are estimated using validated dose-response models, and proton therapy is recommended when the predicted reduction in NTCP exceeds a predefined referral threshold (Langendijk *et al*, 2013; Tambas *et al*, 2020; Langendijk *et al*, 2021; Hansen *et al*, 2024). This provides an individualized and transparent basis for allocating limited proton therapy resources. However, under the current approach, the benefit of normal tissue sparing is evaluated only over the treatment course being planned. This decision horizon may be incomplete when the initial dose distribution also influences treatment options later in the disease course.

Reirradiation represents a clinical setting in which the influence of the primary dose distribution on subsequent treatment options may become particularly relevant. Improvements in cancer detection, systemic therapy and supportive care have increased the number of long-term cancer survivors who may subsequently develop a locoregional recurrence or second primary malignancy within or near a previously irradiated region (Morris *et al*, 2011; Miller *et al*, 2022). Advances in treatment planning and delivery systems have simultaneously expanded the circumstances in which another radiotherapy course can be considered (Romesser *et al*, 2016; Vargo *et al*, 2018; Ward *et al*, 2018; Lee *et al*, 2020). These developments have widened the role of reirradiation, but the dose already delivered to normal tissues during primary treatment remains a key constraint on reirradiation. The inherited dose may determine whether reirradiation can be offered, which prescription dose and target coverage are achievable, and what toxicity risk must be accepted (Chao *et al*, 2017; Dionisi *et al*, 2019; Embring *et al*, 2021). The primary dose distribution therefore determines not only the immediate risk of toxicity but also the therapeutic options available if reirradiation is required.

In this context, upfront proton therapy may have value beyond its immediate reduction in risk of normal tissue complications: by limiting normal tissue exposure during primary treatment, it may also preserve dose capacity for possible future reirradiation. This contingent benefit may be relevant even when the immediate NTCP reduction is insufficient to satisfy the current referral criterion, suggesting that proton therapy allocation should consider both the primary-course benefit and the treatment options preserved if reirradiation becomes necessary. This represents an anticipatory strategy: rather than addressing the inherited dose only when reirradiation is required, it seeks to reduce cumulative normal tissue exposure already when the primary dose landscape is created. Such strategy may complement recent reirradiation research, which has mainly focused on making a second treatment course safer and more effective within the constraints of the inherited primary dose landscape (Andratschke *et al*, 2022; Willmann *et al*, 2023; Paradis *et al*, 2025; Appelt *et al*, 2026; Torelli *et al*, 2026).

To formalize this anticipatory strategy, we developed a reirradiation-aware extension of model-based proton therapy selection. Rather than assessing upfront proton therapy solely according to its predicted benefit after primary treatment, the proposed criterion evaluates its expected benefit across two possible future treatment scenarios: primary treatment alone and primary treatment followed by reirradiation. For the analytically tractable case of a logistic NTCP response, we derived screening conditions from the paired primary photon and proton plans to identify patients for whom future reirradiation could, in principle, justify upfront proton therapy despite a primary-course NTCP benefit below the referral threshold. For patients satisfying these primary-plan conditions, the framework then identifies the reirradiation probabilities and added risk-score contributions under which upfront proton therapy would be justified beyond current selection.

## 2. Model-based proton therapy selection criteria

### 2.1. Current criterion based on primary-treatment outcomes

Current model-based proton therapy selection compares the predicted normal tissue complication probabilities obtained from paired photon and proton plans for the primary treatment (Langendijk *et al*, 2013; Tambas *et al*, 2020). The benefit of selecting proton therapy over photon therapy is then define as

$$\Delta NTCP_{\text{prim}} = NTCP_{\text{prim}}^{X} - NTCP_{\text{prim}}^{p}, \tag{1}$$

with the superscripts $X$ and $p$ denoting photon and proton therapy, respectively. Proton therapy is recommended when this reduction reaches or exceeds a referral threshold $T$:

$$\Delta NTCP_{\text{prim}} \geq T. \tag{2}$$

### 2.2. Reirradiation-aware criterion

Under the current criterion, a patient may achieve meaningful normal tissue sparing with primary proton therapy but still not be referred because the corresponding $\Delta\text{NTCP}_{\text{prim}}$ is small. This may occur, for example, when the primary photon and proton plans lie on a relatively flat region of the response curve (Figure 1a). If the patient subsequently undergoes reirradiation, however, the initial sparing may be carried forward, and the treatment course initiated with protons enters reirradiation with a lower normal tissue burden than the course initiated with photons. If the resulting cumulative treatment effects after reirradiation lie on a steeper region of the response curve, the same primary sparing may translate into a larger and clinically relevant reduction in predicted complication risk (Figure 1b). We denote the benefit conditional on reirradiation as

$$\Delta\text{NTCP}_{\text{reRT}} = \text{NTCP}_{\text{reRT}}^{X} - \text{NTCP}_{\text{reRT}}^{p}. \tag{3}$$

where $\text{NTCP}_{\text{reRT}}^{X}$ and $\text{NTCP}_{\text{reRT}}^{p}$ describe the complication probabilities associated with treatment courses initiated with primary photon and proton therapy, respectively. The outcome model used to estimate these probabilities may differ from the primary-treatment model and need not concern the same complication. To compare and subsequently combine $\Delta\text{NTCP}_{\text{prim}}$ and $\Delta\text{NTCP}_{\text{reRT}}$, however, we assume that both benefit estimates can be expressed on a common, commensurate scale, such that the same numerical reduction represents an equivalent clinical value under either treatment scenario.

To incorporate this potential downstream effect, we propose an extension of the current referral criterion that accounts for the probability of future reirradiation. Let $P_{\text{reRT}}$ denote the probability that the patient will subsequently receive reirradiation, assessed at the time of primary treatment planning. This probability is assumed to be independent of the primary treatment modality. Because it is not yet known which treatment scenario the patient will follow, the expected benefit of selecting proton therapy for the primary course is

$$\mathbb{E}[\Delta\text{NTCP}] = (1 - P_{\text{reRT}})\Delta\text{NTCP}_{\text{prim}} + P_{\text{reRT}}\Delta\text{NTCP}_{\text{reRT}}. \quad (4)$$

With probability $1 - P_{\text{reRT}}$, no reirradiation is delivered and the relevant benefit is $\Delta\text{NTCP}_{\text{prim}}$. With probability $P_{\text{reRT}}$, reirradiation is delivered and the relevant benefit is $\Delta\text{NTCP}_{\text{reRT}}$.

Since this expected benefit is expressed on the same decision scale as $\Delta\text{NTCP}_{\text{prim}}$, the existing referral threshold $T$ can also be applied. Upfront proton therapy is therefore recommended when

$$\mathbb{E}[\Delta\text{NTCP}] \geq T. \quad (5)$$

For $P_{\text{reRT}} = 0$, this reirradiation-aware criterion reduces exactly to the current primary-course criterion. For $P_{\text{reRT}} > 0$, it incorporates the potential downstream value of primary normal tissue sparing. When $\Delta\text{NTCP}_{\text{reRT}} > \Delta\text{NTCP}_{\text{prim}}$, the expected benefit increases with $P_{\text{reRT}}$, potentially justifying proton therapy for a patient who would not otherwise be referred. Conversely, when $\Delta\text{NTCP}_{\text{reRT}} < \Delta\text{NTCP}_{\text{prim}}$, the expected benefit decreases with $P_{\text{reRT}}$, and a patient satisfying the current criterion may fail the reirradiation-aware criterion.

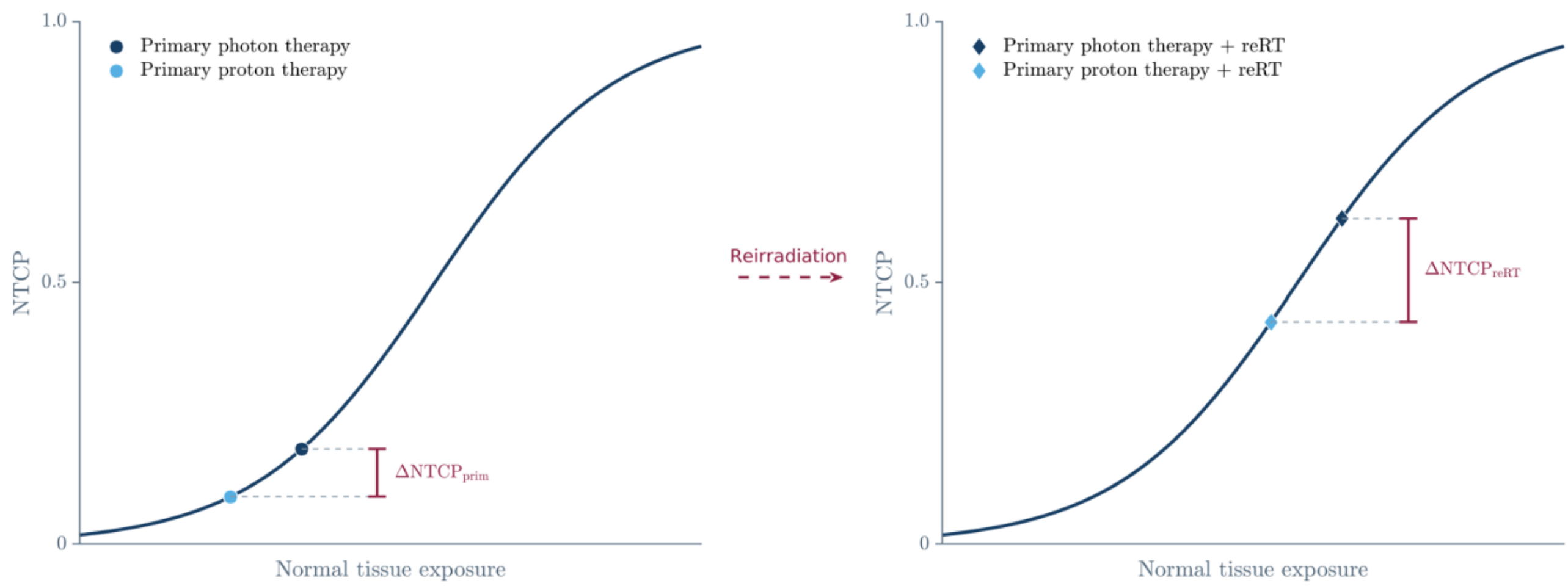

***Figure 1:*** *Schematic illustration of how the benefit of primary proton therapy may differ between the two possible treatment scenarios. A small primary benefit on a flat response region (left) may become larger if the treatment histories move to a steeper region after reirradiation (right). The two scenarios may use different outcome models but must be evaluated on commensurate scales.*

## 3. Analytical characterization of additional referral under a shared logistic NTCP model

The expected-benefit criterion in Eq. (5) determines referral based on the reirradiation probability and the predicted toxicities after reirradiation. At the time of primary treatment planning, however, the future reirradiation scenario is generally uncertain. Prospective use of the reirradiation-aware referral criterion therefore requires answering two consecutive questions:

1. Can the paired primary photon and proton plans identify patients for whom some future reirradiation scenario could, in principle, justify upfront proton therapy?
2. For patients satisfying this initial screening, which combinations of reirradiation probability and added normal tissue exposure would cause the expected benefit to reach the referral threshold?

To answer these questions analytically, we consider the special case in which the complication probabilities after primary treatment and reirradiation can both be represented on a common logistic risk-score scale. The general referral criterion in Section 2 does not require this assumption; it is introduced here to obtain closed-form screening conditions and scenario requirements. The main analysis focuses on patients who do not satisfy the current referral criterion, i.e. for whom

$$\Delta\mathrm{NTCP}_{\mathrm{prim}} < T, \qquad (6)$$

because these patients represent potential additional candidates for upfront proton therapy. The possibility that a patient satisfying the current criterion may fail the reirradiation-aware criterion is considered separately in Appendix A.

### 3.1. Risk-score formulation of primary and reirradiation benefits

On the common logistic scale, the predicted complication probability associated with risk score $S$ is

$$\mathrm{NTCP}(S) = \frac{1}{1+\exp(-S)}. \qquad (7)$$

Here, $S$ is a dimensionless risk score. In simple logistic NTCP models, $S$ is a linear function of a single dosimetric predictor $D$, such as the mean dose to an organ. The analysis could then equivalently be expressed on a dose scale, with $S = 0$ corresponding to $D = TD_{50}$, the dose associated with a 50% complication probability. More generally, the risk score may combine several dosimetric predictors and non-dosimetric clinical factors. We therefore formulate the methodology in terms of $S$ to maintain applicability across different logistic NTCP models.

Let $S^X_{\mathrm{prim}}$ and $S^p_{\mathrm{prim}}$ denote the primary photon and proton risk scores, respectively. Their separation is defined as

$$\Delta S = S^X_{\mathrm{prim}} - S^p_{\mathrm{prim}}. \qquad (8)$$

The primary-treatment benefit can therefore be described as

$$\Delta\mathrm{NTCP}_{\mathrm{prim}} = \frac{1}{1+\exp\left(-S_{\mathrm{prim}}^{X}\right)} - \frac{1}{1+\exp\left[-\left(S_{\mathrm{prim}}^{X} - \Delta S\right)\right]}. \tag{9}$$

Possible future reirradiation is represented by a non-negative added risk-score contribution $S_{\mathrm{reRT}}$. For the present analytical specialization, the same contribution is applied after either primary modality. The risk scores conditional on reirradiation are consequently

$$S_{\mathrm{cum}}^{X} = S_{\mathrm{prim}}^{X} + S_{\mathrm{reRT}} \tag{10}$$

and

$$S_{\mathrm{cum}}^{p} = S_{\mathrm{prim}}^{X} - \Delta S + S_{\mathrm{reRT}}. \tag{11}$$

This assumption preserves the photon-proton score separation $\Delta S$ while changing the position of both treatment courses on the logistic response curve. The cumulative benefit conditional on reirradiation is then

$$\Delta\mathrm{NTCP}_{\mathrm{reRT}} = \frac{1}{1+\exp\left[-\left(S_{\mathrm{prim}}^{X} + S_{\mathrm{reRT}}\right)\right]} - \frac{1}{1+\exp\left[-\left(S_{\mathrm{prim}}^{X} - \Delta S + S_{\mathrm{reRT}}\right)\right]}. \tag{12}$$

Using the hyperbolic representation of the logistic function, Eq. (12) can equivalently be written as

$$\Delta\mathrm{NTCP}_{\mathrm{reRT}} = \frac{\sinh(\Delta S/2)}{\cosh\left(S_{\mathrm{prim}}^{X} + S_{\mathrm{reRT}} - \Delta S/2\right) + \cosh(\Delta S/2)}. \tag{13}$$

Equation (13) makes explicit the two quantities governing the reirradiation-course benefit. The score separation $\Delta S$ determines the largest NTCP reduction that can be produced, whereas $S_{\mathrm{prim}}^{X} + S_{\mathrm{reRT}}$ determines where this separation is positioned on the logistic response curve. For a fixed $\Delta S$, the numerator is constant and the denominator is minimized when

$$S_{\mathrm{prim}}^{X} + S_{\mathrm{reRT}} = \frac{\Delta S}{2}. \tag{14}$$

At this position, the photon and proton scores lie symmetrically around the inflection point of the logistic response and their NTCP difference is maximal. Moving the two scores towards this position increases $\Delta\mathrm{NTCP}_{\mathrm{reRT}}$, whereas moving them beyond it decreases the benefit as both predicted probabilities approach saturation. These properties provide the basis for the screening conditions and scenario requirements below.

### 3.2. Screening conditions based on primary-treatment information

The first stage asks whether any combination of reirradiation probability $P_{\mathrm{reRT}} \in [0, 1]$ and non-negative score contribution $S_{\mathrm{reRT}}$ could allow a patient satisfying Eq. (6) to meet the proposed reirradiation-aware criterion. Because the expected benefit is a weighted average of $\Delta\mathrm{NTCP}_{\mathrm{prim}}$ and $\Delta\mathrm{NTCP}_{\mathrm{reRT}}$, and because we assumed $\Delta\mathrm{NTCP}_{\mathrm{prim}} < T$, additional referral is possible only if the reirradiation-course benefit can reach at least the referral threshold:

$$\Delta\mathrm{NTCP}_{\mathrm{reRT}} \geq T \tag{15}$$

under some admissible value of $S_{\mathrm{reRT}}$. If both scenario-specific benefits remain below $T$, their expected value cannot reach the threshold for any reirradiation probability. The existence of a score contribution satisfying Eq. (15) is determined by two conditions that depend only on the paired primary scores and the referral threshold.

**Condition 1 (sufficient photon-proton score separation):** For a fixed $\Delta S$, the largest possible NTCP reduction is obtained at the symmetric position in Eq. (14). Substitution of this position into Eq. (13) gives

$$\Delta\mathrm{NTCP}_{\mathrm{reRT,max}} = \tanh\left(\frac{\Delta S}{4}\right). \tag{16}$$

For a referral-level reirradiation-course benefit to be possible, this maximum must satisfy

$$\tanh\left(\frac{\Delta S}{4}\right) \geq T, \tag{17}$$

which is equivalent to

$$\Delta S \geq 4\,\mathrm{arctanh}(T). \tag{18}$$

If Condition 1 is not satisfied, the photon-proton score separation is insufficient to produce an NTCP reduction of $T$, irrespective of the reirradiation probability, the primary score position or the added score contribution.

**Condition 2 (reachability of the benefit-producing region):** When Condition 1 is satisfied, solving Eq. (13) at $\Delta\mathrm{NTCP}_{\mathrm{reRT}} = T$ gives the lower and upper cumulative photon scores between which the reirradiation-course benefit reaches the referral threshold:

$$S^{X}_{\mathrm{cum},\pm}(T) = \frac{\Delta S}{2} \pm \mathrm{arcosh}\left[\frac{\sinh(\Delta S/2)}{T} - \cosh(\Delta S/2)\right]. \tag{19}$$

These cumulative photon risk scores lie symmetrically around the position of maximum benefit, $S^{X}_{\mathrm{cum}} = \Delta S/2$. For the given score separation $\Delta S$, the reirradiation-course benefit satisfies $\Delta\mathrm{NTCP}_{\mathrm{reRT}} \geq T$ whenever

$$S^{X}_{\mathrm{cum},-}(T) \leq S^{X}_{\mathrm{cum}} \leq S^{X}_{\mathrm{cum},+}(T). \tag{20}$$

Because patients considered in the main analysis satisfy Eq. (6), their primary photon score lies outside this interval, either below its lower boundary or above its upper boundary. Furthermore, because $S_{\mathrm{reRT}} \geq 0$, reirradiation can only increase the photon score according to $S^{X}_{\mathrm{cum}} = S^{X}_{\mathrm{prim}} + S_{\mathrm{reRT}}$. The benefit-producing interval is therefore reachable if and only if

$$S^{X}_{\mathrm{prim}} < S^{X}_{\mathrm{cum},+}(T) \tag{21}$$

A positive $S_{\mathrm{reRT}}$ can then move both treatment courses into the benefit-producing region. If the primary photon score instead lies above $S^{X}_{\mathrm{cum},+}(T)$, any further increase moves both predicted probabilities farther into the saturated part of the logistic response, and the required benefit cannot be reached.

Conditions 1 and 2 are jointly necessary and sufficient for the existence of at least one non-negative reirradiation score contribution for which $\Delta\mathrm{NTCP}_{\mathrm{reRT}} \geq T$. They can therefore be evaluated using only the paired primary plan scores and the referral threshold, without specifying a future reirradiation probability or score contribution.

### 3.3. Reirradiation probability and risk score contribution requirements

Passing the screening conditions establishes that additional referral is mathematically possible, but not that it occurs under a clinically plausible reirradiation scenario. The second stage therefore quantifies two complementary requirements: the minimum reirradiation probability under the most favorable attainable score contribution and, for an assumed probability, the range of score contributions for which the expected benefit reaches the referral threshold.

**Minimum reirradiation probability:** For an initially non-referred patient satisfying both screening conditions, $S_{\mathrm{prim}}^{X}$ lies below the benefit-producing region and therefore below the position of maximum benefit in Eq. (14). This position can be reached using the non-negative contribution

$$S_{\mathrm{reRT}}^{*} = \frac{\Delta S}{2} - S_{\mathrm{prim}}^{X}. \quad (22)$$

The maximum attainable reirradiation-course benefit is consequently the value in Eq. (16). Substituting this maximum into the expected-benefit criterion in Eq. (5) and solving for $P_{\mathrm{reRT}}$ gives the minimum reirradiation probability required for additional referral:

$$P_{\mathrm{reRT,min}} = \frac{T - \Delta\mathrm{NTCP}_{\mathrm{prim}}}{\tanh(\Delta S/4) - \Delta\mathrm{NTCP}_{\mathrm{prim}}}. \quad (23)$$

For profiles satisfying both screening conditions, $P_{\mathrm{reRT,min}} \in (0,1]$. If the estimated reirradiation probability is below this value, the expected benefit cannot reach the referral threshold for any non-negative value of $S_{\mathrm{reRT}}$.

**Admissible reirradiation score-contribution window:** Conversely, for an assumed probability $P_{\mathrm{reRT}} > 0$, rearrangement of the expected-benefit criterion gives the minimum reirradiation-course benefit required for referral:

$$\Delta\mathrm{NTCP}_{\mathrm{reRT,min}} = \Delta\mathrm{NTCP}_{\mathrm{prim}} + \frac{T - \Delta\mathrm{NTCP}_{\mathrm{prim}}}{P_{\mathrm{reRT}}}. \quad (24)$$

An admissible score contribution exists only when

$$P_{\mathrm{reRT}} \geq P_{\mathrm{reRT,min}}, \quad (25)$$

such that the required benefit in Eq. (24) does not exceed the maximum attainable value in Eq. (16).

Replacing $T$ in Eq. (19) with $\Delta\mathrm{NTCP}_{\mathrm{reRT,min}}$ gives the cumulative photon-score boundaries within which the required benefit is produced:

$$S_{\mathrm{cum},\pm}^{X} = \frac{\Delta S}{2} \pm \operatorname{arcosh}\left[\frac{\sinh(\Delta S/2)}{\Delta \mathrm{NTCP}_{\mathrm{reRT,min}}} - \cosh(\Delta S/2)\right]. \quad (26)$$

Because $S_{\mathrm{cum}}^{X} = S_{\mathrm{prim}}^{X} + S_{\mathrm{reRT}}$, the admissible reirradiation score contributions satisfy

$$\max\{0,\, S_{\mathrm{cum},-}^{X} - S_{\mathrm{prim}}^{X}\} \le S_{\mathrm{reRT}} \le S_{\mathrm{cum},+}^{X} - S_{\mathrm{prim}}^{X}. \quad (27)$$

At $P_{\mathrm{reRT}} = P_{\mathrm{reRT,min}}$, the lower and upper boundaries coincide at the score contribution in Eq. (22). As the assumed reirradiation probability increases, a smaller conditional benefit is required and the admissible interval broadens. The interval nevertheless remains bounded because excessively large values of $S_{\mathrm{reRT}}$ move both predicted risks into the saturated part of the logistic response and reduce their difference.

Together, the two-stage analysis separates information available from the primary plans from assumptions about the uncertain future course. The primary photon score $S_{\mathrm{prim}}^{X}$, the photon-proton score separation $\Delta S$, and the threshold $T$ first determine whether additional referral is possible. For profiles satisfying both screening conditions, $P_{\mathrm{reRT,min}}$ and the admissible $S_{\mathrm{reRT}}$ interval then identify the reirradiation scenarios under which upfront proton therapy may satisfy the proposed criterion.

## 4. Characterization of referral patterns in the logistic risk-score space

### 4.1. Evaluation procedure

To characterize the behavior of the reirradiation-aware criterion and its implications for proton therapy referral, we evaluated it directly in the logistic risk-score space. This formulation isolates the effects of the primary photon score $S_{\mathrm{prim}}^{X}$, the photon-proton score separation $\Delta S$, the reirradiation probability $P_{\mathrm{reRT}}$ and the added risk-score contribution $S_{\mathrm{reRT}}$, without relying on a particular disease site, dose metric or toxicity model. The evaluation was deterministic and no population distribution was assigned to the risk-score profiles. Consequently, the analysis characterizes where referral and additional referral can occur, but does not estimate the proportions of patients belonging to the respective regions in a clinical population. An illustrative referral threshold of $T = 0.10$ was used throughout the numerical evaluation.

We first examined three representative profiles to illustrate separately how the primary score position and the photon-proton score separation influence the expected NTCP benefit. A reference profile A was compared with a second profile B having the same primary photon score but a smaller $\Delta S$, thereby isolating the effect of reduced photon-proton separation. The third profile C retained the same $\Delta S$ as the reference profile A but had a lower primary photon score, allowing the effect of score position on the logistic response to be examined independently. The selected profiles are summarized in Table 1. For each profile, the expected benefit was evaluated as a function of $P_{\mathrm{reRT}}$ for several fixed values of $S_{\mathrm{reRT}}$. This analysis shows whether and at which reirradiation probability the expected benefit reaches the referral threshold, and how this transition changes when either the primary score position or the photon-proton score separation is modified.

***Table 1:*** *Primary risk-score characteristics and screening outcomes for the three illustrative profiles. All profiles have $\Delta NTCP_{prim} < T$ and are therefore not referred under the current criterion, but satisfy both necessary conditions for possible additional referral under the reirradiation-aware criterion. Values are reported for $T = 0.10$.*

| **Patient profile** | $S_{\mathrm{prim}}^{X}$ | $\Delta S$ | $\Delta\mathrm{NTCP}_{\mathrm{prim}}$ | **Condition 1** | **Condition 2** |
|---|---|---|---|---|---|
| A | -1.5 | 0.8 | 0.091 | Satisfied | Satisfied |
| B | -1.5 | 0.7 | 0.083 | Satisfied | Satisfied |
| C | -2.0 | 0.7 | 0.062 | Satisfied | Satisfied |

We then extended the evaluation to the full two-dimensional risk-score space. Multiple combinations of $S_{\mathrm{prim}}^{X}$ and $\Delta S$ were evaluated on a dense grid under different reirradiation scenarios, each defined by a combination of $P_{\mathrm{reRT}}$ and $S_{\mathrm{reRT}}$. At each grid point, we calculated $\Delta\mathrm{NTCP}_{\mathrm{prim}}$, $\Delta\mathrm{NTCP}_{\mathrm{reRT}}$, and the expected NTCP benefit in Eq. (4). Profiles were classified as referrals under the current criterion, additional referrals under the reirradiation-aware criterion, or non-referrals under either criterion. We additionally identified the regions excluded from additional referral because they failed Condition 1, owing to insufficient photon-proton score separation, or Condition 2, because the primary photon score lay beyond the benefit-producing region reachable through a non-negative $S_{\mathrm{reRT}}$. Reclassification of patients already satisfying the current criterion was considered separately in Appendix A.

### 4.2. Evaluation results

#### 4.2.1. Influence of primary score position and photon-proton separation

Figure 2 illustrates how the primary photon score and photon-proton score separation jointly determine the response to possible reirradiation. All three profiles had $\Delta\mathrm{NTCP}_{\mathrm{prim}} < 0.10$ and therefore failed the current referral criterion for the assumed threshold, although each satisfied both screening conditions. Consequently, additional referral was possible for all three profiles, but depended on the assumed $S_{\mathrm{reRT}}$ and $P_{\mathrm{reRT}}$.

For the reference profile A, $S_{\mathrm{reRT}} = 2$ moved the photon and proton scores towards the steepest region of the logistic response, increasing the conditional benefit from $\Delta\mathrm{NTCP}_{\mathrm{prim}} = 0.091$ to $\Delta\mathrm{NTCP}_{\mathrm{reRT}} = 0.197$. The expected benefit reached the referral threshold at $P_{\mathrm{reRT}} = 0.082$. With $S_{\mathrm{reRT}} = 3$, the scores were shifted further along the response curve, resulting in a smaller $\Delta\mathrm{NTCP}_{\mathrm{reRT}}$ of 0.149 and a higher required probability of 0.150. By contrast, $S_{\mathrm{reRT}} = 4$ moved both scores into the upper, saturating region of the curve. The conditional benefit then decreased to 0.079, below the primary benefit, and the expected benefit declined with increasing $P_{\mathrm{reRT}}$. Profile B had the same primary photon score as profile A but a smaller photon-proton score separation. This reduced both the primary benefit and the maximum benefit generated after reirradiation. The referral threshold was reached at $P_{\mathrm{reRT}} = 0.193$ for $S_{\mathrm{reRT}} = 2$ and at $P_{\mathrm{reRT}} = 0.386$ for $S_{\mathrm{reRT}} = 3$. As for profile A, $S_{\mathrm{reRT}} = 4$ shifted both scores beyond the benefit-producing region, so that no additional referral occurred. The comparison between profiles A and B therefore shows that, for a similar initial position on the response curve, a larger photon-proton score separation produces a larger potential benefit and lowers the reirradiation probability required for referral. Profile C had the same score separation as profile A but a lower primary photon score. Its primary benefit was therefore smaller because both primary plans were positioned farther into the lower, flatter region of the logistic response. Contributions of $S_{\mathrm{reRT}} = 2$ and $S_{\mathrm{reRT}} = 3$ moved the scores into the steeper part of the curve and produced similar conditional benefits of 0.190 and 0.181, respectively. The referral threshold was reached at probabilities of 0.298 and 0.319. Because of its lower starting position, profile C approached saturation later than profiles A and B: even with $S_{\mathrm{reRT}} = 4$, its conditional benefit remained above the referral threshold, although additional referral required a comparatively high probability of 0.756.

Together, these profiles demonstrate the distinct roles of the two primary-plan characteristics. The score separation determines the magnitude of the benefit that can potentially be produced, whereas the primary photon score determines which part of the logistic response is reached for a given $S_{\mathrm{reRT}}$. Consequently, increasing $S_{\mathrm{reRT}}$ does not necessarily increase the value of upfront proton therapy: the expected benefit increases as the cumulative scores approach the steep region of the response curve but decreases once they move towards saturation.

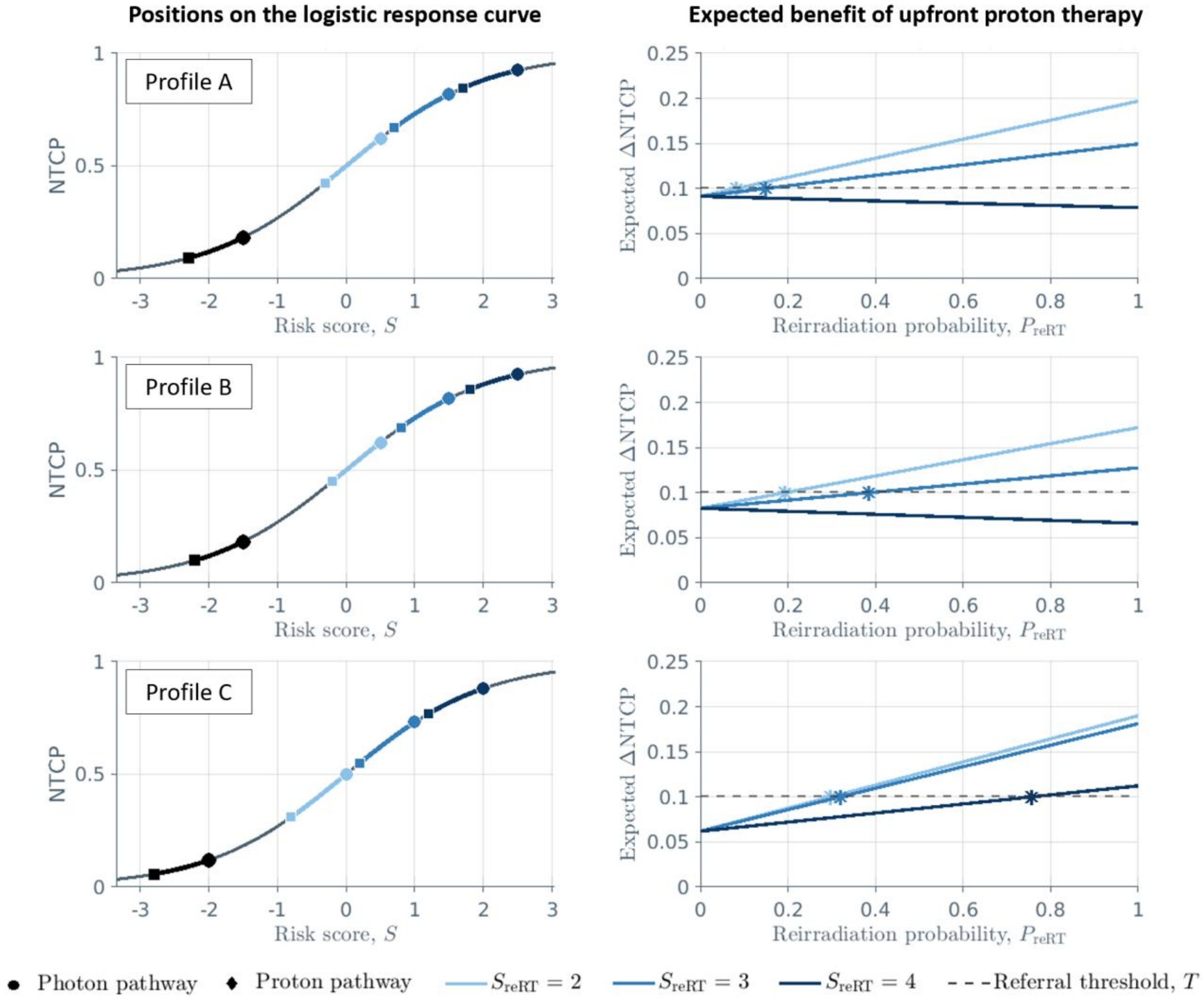

***Figure 2:*** *Effects of primary score position and photon-proton score separation on reirradiation-aware referral. The three rows show profiles A-C; the left panels show their positions on the logistic response curve for* $S_{reRT} = 2, 3, 4$*, and the right panels show the corresponding expected benefit as a function of* $P_{reRT}$*. Asterisks mark threshold crossings.*

### 4.2.2. Referral regions under illustrative reirradiation scenarios

Figure 3 maps the referral decision as a function of the primary photon score $S_{\text{prim}}^X$ and the photon-proton score separation $\Delta S$, for reirradiation probabilities $P_{\text{reRT}} \in \{0.10, 0.20\}$ and reirradiation score contributions $S_{\text{reRT}} \in \{2, 3, 4\}$. The dark blue regions represent patients satisfying the current referral criterion, whereas the red regions identify additional patients who do not currently qualify but reach the expected-benefit threshold after accounting for possible reirradiation. The remaining unhatched grey regions satisfy both analytical screening conditions but do not qualify for referral under the corresponding scenario.

The screening conditions exclude two distinct parts of the decision space before a specific reirradiation scenario is evaluated. The horizontal hatched region at small $\Delta S$ fails Condition 1 because the preserved score separation is insufficient to produce an NTCP benefit of magnitude $T$, even at its theoretically optimal operating point. Consequently, this region also lies entirely outside the current referral region. The second hatched region passes Condition 1 but fails Condition 2: although the score separation is sufficiently large, the primary operating point lies beyond the range from which the required benefit can be reached following reirradiation. Passing both conditions therefore identifies patients for whom additional referral may be possible, but does not itself establish referral.

The current referral region is independent of $P_{\mathrm{reRT}}$ and $S_{\mathrm{reRT}}$, and is consequently identical across all six panels. In contrast, the additional referral region depends strongly on both quantities. Increasing $P_{\mathrm{reRT}}$ from 0.10 to 0.20 increases potential referral to proton therapy because a smaller cumulative benefit is sufficient to raise the expected benefit above $T$. Increasing $S_{\mathrm{reRT}}$ shifts the additional referral region towards more negative values of $S_{\mathrm{prim}}^{X}$. Patients starting further along the low-risk tail of the response curve for the primary treatment can then be moved into the steeper part for reirradiation, where the score separation between photon and proton therapy translates into a larger NTCP difference. Conversely, for patients with less negative primary scores, a larger $S_{\mathrm{reRT}}$ can shift both cumulative scores towards saturation, limiting the additional benefit.

These results demonstrate that $\Delta S$ alone is insufficient to determine whether reirradiation considerations alter the referral decision. Patients with the same photon-proton score separation can have markedly different referral outcomes depending on their primary photon score and the anticipated reirradiation scenario. Additional referral is therefore concentrated among patients who combine a sufficiently large preserved score separation with a primary operating point that can be shifted into a benefit-producing region of the response curve by a plausible future reirradiation course.

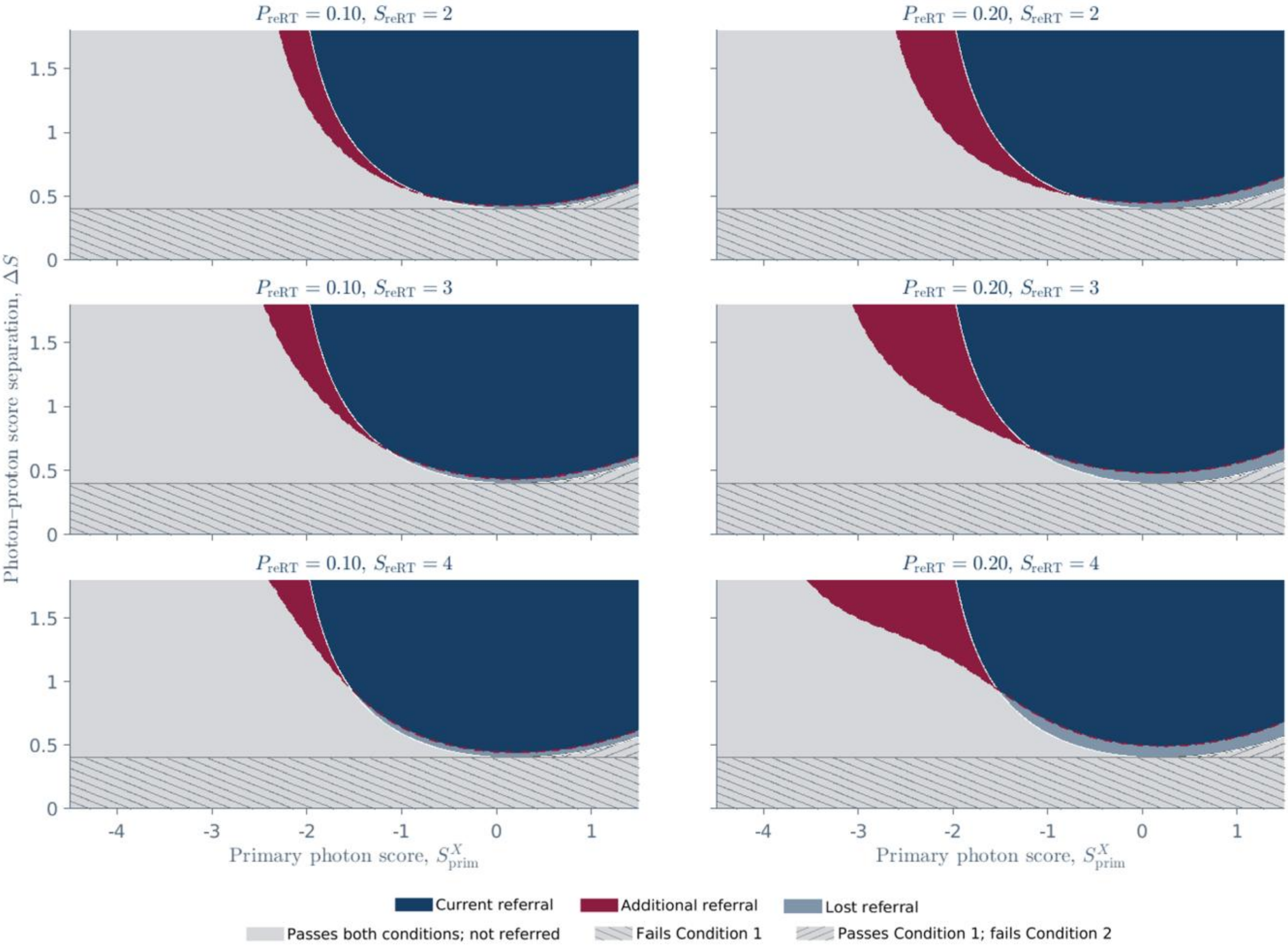

***Figure 3:*** *Referral regions in logistic risk-score space for different $P_{reRT}$ and $S_{reRT}$. Dark blue and red indicate current and additional referral, respectively, for a referral threshold $T$ of 0.10; grey indicates no referral, with hatching identifying profiles excluded by the analytical screening conditions. Light blue regions instead indicate patients referred according to the current criterion, but which may lose indication for upfront proton therapy based on the reirradiation-aware referral.*

## 5. Discussion

This study introduces a reirradiation-aware extension of model-based proton therapy selection that assigns value to primary normal tissue sparing according to both its immediate benefit and its potential relevance if reirradiation is subsequently required. The analytical results show that additional referral could be justified only for a restricted subset of patients with well-defined paired photon-proton plan characteristics. These profiles require sufficient photon-proton score separation and a primary score position from which additional exposure can move the treatment histories into a region of the response curve producing a greater NTCP separation. Importantly, the framework separates information available from the primary plans from assumptions about the uncertain future course. The two screening conditions first establish whether additional referral is possible in principle; only for profiles satisfying them must the reirradiation probability and added risk-score contribution be considered to identify scenarios supporting upfront proton therapy.

This anticipatory perspective complements current research on reirradiation, which aims to reduce cumulative toxicity while generally treating the primary dose distribution as fixed (Andratschke *et al*, 2022; Beddok *et al*, 2025). Existing work has focused on selecting patients likely to benefit from another local treatment (Ward *et al*, 2018, Ward *et al*, 2019, Lee *et al*, 2020), reconstructing cumulative normal tissue exposure through biological dose conversion and spatial dose registration (Brock *et al*, 2017; Paradis *et al*, 2019; Appelt *et al*, 2026), characterizing tissue recovery (Nieder *et al*, 2006; Dörr and Gabryś 2018) and cumulative dose-toxicity relationships (Chao *et al*, 2017; Embring *et al*, 2021), and optimizing the beam arrangement (Torelli *et al*, 2025) and prescription dose of the second course within the remaining therapeutic window (Caudell *et al*, 2018). Proton therapy is also commonly considered at this stage because its dose-sparing properties may facilitate reirradiation while limiting further normal tissue exposure (Romesser *et al*, 2016; Phan *et al*, 2016; Chao *et al*, 2017; Lee *et al*, 2023). The proposed strategy moves this decision upstream by asking whether proton therapy should instead be selected during primary treatment to reduce the constraints inherited by a possible future radiation course. To our knowledge, this is the first formal framework to incorporate anticipated reirradiation into model-based selection of the primary treatment modality.

The potential value of this strategy may extend beyond the reduction in expected complication probability. Normal tissue exposure avoided during primary treatment could influence whether reirradiation remains feasible, which prescription dose and target coverage can be achieved, and whether treatment with curative intent remains possible. Patient selection should therefore reflect not only the probability of recurrence, but also the probability and expected clinical value of receiving reirradiation. Patients with aggressive disease, early or disseminated relapse, poor performance status or substantial baseline organ dysfunction may have a high probability of recurrence but little opportunity to benefit from aggressive salvage treatment (Ward *et al*, 2018; Lee *et al*, 2020; Lee *et al*, 2023). Conversely, patients with a longer life expectancy and a greater likelihood of an isolated, salvageable recurrence may benefit from preserved normal tissue capacity despite having a lower recurrence risk (Ward *et al*, 2018; Lee *et al*, 2020). HPV-associated head-and-neck cancer illustrates this distinction: selected patients with HPV-positive disease may have a comparatively favorable prognosis and greater potential to benefit from curative salvage, whereas a higher recurrence probability in HPV-negative disease does not necessarily imply greater reirradiation value (Davis *et al*, 2014; Fakhry *et al*, 2014; Guo *et al*, 2015; Christopherson *et al*, 2021). Similar considerations may apply to long-term survivors at risk of an isolated locoregional recurrence or second primary malignancy (Morris *et al*, 2011; Miller *et al*, 2022). Other relevant factors therefore include survival, recurrence pattern, disease-free interval, resection status, alternative salvage options, treatment intent, competing risks and patient preferences.

Several limitations must be considered. Most importantly, no NTCP model based on cumulative exposure after reirradiation has yet been clinically validated for this purpose. Clinical translation therefore requires validated outcome models for the reirradiation scenario and a defensible common scale on which benefits across the two possible treatment courses can be compared. The primary and reirradiation outcomes need not concern the same complication, but equal numerical reductions do not necessarily represent equivalent clinical value; differences in severity, duration, reversibility and quality-of-life impact may require utility weighting. The logistic analysis additionally assumes a common added risk-score contribution and retention of the primary photon-proton score separation, whereas actual outcomes may depend on spatial dose accumulation, fractionation, treatment interval, normal tissue recovery, anatomical changes and optimization of the subsequent plan. Moreover, the probability of reirradiation is assumed to be independent of the primary modality, although greater preservation of normal tissue capacity could itself influence whether another radiotherapy course can be offered. The deterministic score-space analysis identifies where additional referral can occur but does not estimate the prevalence of these profiles in a clinical population.

The proposed strategy could also increase the use of limited proton capacity during primary treatment. Although upfront proton therapy might preserve sufficient normal tissue capacity to permit a later reirradiation course with photons and thereby reduce subsequent reliance on proton therapy, such a capacity benefit is not guaranteed. It will depend on how accurately patients who will eventually receive clinically meaningful reirradiation can be identified and on the relative resource requirements of primary and salvage treatment. Furthermore, applying the expected-benefit criterion symmetrically could remove some patients who satisfy the current criterion when their predicted benefit after reirradiation is smaller. A conservative implementation could instead preserve established referrals and apply the proposed framework only to additional candidates.

Future work should evaluate the framework in longitudinal, disease-specific cohorts containing paired primary plans, recurrence patterns, salvage treatments and long-term outcomes. Such studies should estimate the probability of receiving clinically meaningful reirradiation rather than recurrence alone and develop reirradiation outcome models incorporating cumulative exposure, treatment interval and relevant prognostic factors. The common decision scale should account for toxicity severity, quality of life, survival and treatment feasibility. Ultimately, clinical studies and health-economic analyses will be needed to determine whether anticipatory selection improves tumor control and patient-reported outcomes while using limited proton therapy capacity efficiently.

## 6. Conclusion

We developed a reirradiation-aware extension of model-based proton therapy selection that considers both the immediate benefit of primary normal tissue sparing and its potential value if reirradiation is subsequently required. The proposed framework identifies a restricted group of patient profiles for whom primary proton therapy may be justified beyond the current referral strategy. These profiles are characterized by sufficient photon-proton score separation but a small primary benefit because the plans lie in a less responsive region of the dose-response curve. After reirradiation, however, the retained sparing may become more clinically relevant. Clinical application will require validated reirradiation outcome models, individualized estimates of the probability and clinical value of reirradiation, and prospective evaluation of patient outcomes and proton capacity use.

## Appendix A. Potential loss of referral under the reirradiation-aware criterion

The main analysis considers patients who do not satisfy the current primary-course criterion and evaluates whether the proposed framework identifies additional candidates for proton therapy. If applied symmetrically to all patients, however, the expected-benefit criterion could also change the eligibility of patients who are currently referred. These patients satisfy

$$\Delta\mathrm{NTCP}_{\mathrm{prim}} \geq T. \tag{A.1}$$

Using Eq. (4), their expected benefit can be re-formulated as

$$\mathbb{E}[\Delta\mathrm{NTCP}] = \Delta\mathrm{NTCP}_{\mathrm{prim}} - P_{\mathrm{reRT}}\left(\Delta\mathrm{NTCP}_{\mathrm{prim}} - \Delta\mathrm{NTCP}_{\mathrm{reRT}}\right). \tag{A.2}$$

Equation (A.2) shows that referral is preserved whenever $\Delta\mathrm{NTCP}_{\mathrm{reRT}} \geq \Delta\mathrm{NTCP}_{\mathrm{prim}}$, because the expected benefit cannot fall below the primary-course benefit. When $\Delta\mathrm{NTCP}_{\mathrm{reRT}} < \Delta\mathrm{NTCP}_{\mathrm{prim}}$, however, the expected benefit decreases with increasing reirradiation probability. In this case, the patient fails the reirradiation-aware criterion when

$$P_{\mathrm{reRT}} > \frac{\Delta\mathrm{NTCP}_{\mathrm{prim}} - T}{\Delta\mathrm{NTCP}_{\mathrm{prim}} - \Delta\mathrm{NTCP}_{\mathrm{reRT}}}. \tag{A.3}$$

Such a probability exists within $P_{\mathrm{reRT}} \in [0, 1]$ only if

$$\Delta\mathrm{NTCP}_{\mathrm{reRT}} < T. \tag{A.4}$$

Thus, a currently referred patient can lose referral only when the benefit conditional on reirradiation is below the referral threshold and reirradiation is assigned sufficient probability. For a patient whose primary-course benefit lies exactly at the threshold, any positive reirradiation probability leads to loss of referral if $\Delta\mathrm{NTCP}_{\mathrm{reRT}} < T$.

This possibility is an inherent consequence of applying the expected-benefit criterion symmetrically: a possible future course can decrease, as well as increase, the value assigned to primary proton therapy. Clinically, however, this would withdraw an indication supported by the established primary-treatment criterion on the basis of an uncertain future event and a reirradiation outcome model. Moreover, clinical selection protocols may combine several endpoint-specific criteria, such that failure of one criterion would not necessarily imply overall exclusion from proton therapy. A conservative implementation could therefore preserve current referrals and apply the reirradiation-aware criterion only to patients who are not otherwise eligible. Under this asymmetric policy, the proposed framework identifies additional candidates without removing established referrals; using it for de-referral would require separate clinical and policy justification.